\documentstyle[preprint,aps]{revtex}

\begin{document}

\newcommand\beq{\begin{equation}}
\newcommand\eeq{\end{equation}}
\newcommand\bea{\begin{eqnarray}}
\newcommand\eea{\end{eqnarray}}

\draft

\title{Novel correlations in two dimensions: Some exact solutions}
\author{M. V. N. Murthy\thanks{Permanent Address: The 
Institute of Mathematical Sciences, Madras 600 113, India},
R. K. Bhaduri and Diptiman Sen\thanks{Permanent Address: Indian Institute
of Science, Bangalore 560012, India}}
\address
{Department of Physics and Astronomy, McMaster University, \\
Hamilton, Ontario, Canada L8S 4M1} 
\date{\today}
\maketitle
\begin{abstract}

We construct a new many-body Hamiltonian with two- and three-body interactions 
in two space dimensions and obtain its exact many-body ground state for an 
arbitrary number of particles. This ground state has a novel pairwise 
correlation. A class of exact solutions for the excited states is also found. 
These excited states display an energy spectrum similar to the 
Calogero-Sutherland model in one dimension.  The model reduces to an analog 
of the well-known trigonometric Sutherland model when projected on to a 
circular ring. 

\end{abstract} 
\vskip 1 true cm

\narrowtext

There has been a renewed interest in exactly solvable many-body problems
in recent times. A celeberated example of a solvable many-body problem is
the well-known Calogero-Sutherland model (CSM) in one 
dimension \cite{C69,S71,S72}. The particles in the CSM are confined in a
one-body oscillator potential or on the rim of a circle, and interact with each 
other through a two-body potential which varies as the inverse-square of the
distance between particles. The CSM and its variants in one-dimension, like
the Haldane-Shastry model for spin chains \cite{H88}, have provided
the paradigms to analyze more complicated interacting systems. It is
therefore of great interest to find analogous models in higher dimensions
which share features similar to the CSM. A characteristic feature of the
CSM is the structure of the highly correlated wave function. The
correlations are built into the exact wave function through a Jastrow
factor $(x_i -x_j)^{\lambda} |x_i-x_j|^{\alpha}$ for any pair of particles
denoted by $i,j$. The exponents on the correlator are related to the
strength of the inverse-square interaction. Notice that this factor is
antisymmetric (symmetric) in particle labels for $\lambda=1(0)$ and
vanishes as the two particles approach each other. A generalization of
this in two dimensions is to be found in the Laughlin's trial wave
function \cite{L83} where the correlations are built in through the factor
$(z_i - z_j)$, where $z_i$ are the particle coordinates in complex notation.
The corresponding Hamiltonian for which the Laughlin wave function is 
an exact ground state has not been analyzed to the same degree of detail
as the CSM. However it is known that it is the ground state for a
Hamiltonian describing spin polarized electrons in the lowest Landau level
with a short-range repulsive interaction \cite{TK85}. 

In two dimensions, however, there exists another form of the pair correlator 
with which a Jastrow-type many-body wave function may be constructed, namely 
\beq
X_{ij} ~=~ x_i y_j ~-~ x_j y_i ~.  
\label{xij} 
\eeq
This type of correlation is by definition antisymmetric and goes to zero
as two particles approach each other. In addition, it introduces 
zeros in the wave function whenever the relative angle between the two 
particles goes to zero or $\pi$. The difference with the Jastrow-Laughlin 
form is also significant; (\ref{xij}) is a pseudoscalar. Unlike the 
Laughlin type correlation, it does not impart any 
angular momentum to the two dimensional wavefunction. We now ask: Is
there a Hamiltonian for which the exact solutions have this type of
correlation? In this paper, we construct a many-body Hamiltonian whose
exact ground state involves a product of the pairwise correlations of the
form given by Eq. (\ref{xij}). While we are able to find a class of excited
states of the Hamiltonian, we have not been able to solve for all the
excited states. The known spectrum of states bears a remarkable similarity to
the spectrum of the CSM in one dimension; namely, the spectrum (which includes
the ground state and part of the excited states) is bodily shifted by an 
$N$-dependent but configuration 
independent factor. This is the feature which is responsible for ideal 
fractional exclusion statistics \cite{H91,W94} in the CSM \cite{MS94}. 

The $N-$particle Hamiltonian which we consider is given by
\beq
H~=~ -{\hbar^2\over {2m}} ~\sum_{i=1}^{N} {\vec \nabla}^2_i + {{m\omega^2}
\over 2} ~\sum_{i=1}^{N} {\vec r}_i^2 + \frac{\hbar^2}{2m} g_1 ~
\sum_{\stackrel{i,j}{ (i\ne j)}}^{N} \frac{{\vec r}_j^2}{X_{ij}^2}
+ \frac{\hbar^2}{2m} g_2 ~\sum_{\stackrel{i,j,k}{ (i\ne j\ne k)}}^{N} 
\frac{{\vec r}_j \cdot {\vec r}_k}{X_{ij}X_{ik}} ~,    
\eeq
where $X_{ij}$ is given by (\ref{xij}); $g_1$ and $g_2$ are
dimensionless coupling strengths of the two- and three-body interactions
respectively. While $g_1$ and $g_2$ can be independent of each 
other in general, for the type of solutions involving the correlator in 
(\ref{xij}) they are not. We will specify their relationship shortly. The 
particles are confined in a one-body oscillator
confinement potential. The Hamiltonian is manifestly rotationally
invariant and symmetric in all particle indices. As in the CSM, we may scale
away the mass $m$ and oscillator frequency $\omega$ by scaling all distances
$\vec r_i \rightarrow \sqrt{m\omega/\hbar}~ \vec r_i$, and measuring the
energy in units of $\hbar \omega$. This is done by simply setting
$\hbar =m=\omega =1$. In these units the Hamiltonian is given by
\beq
H~=~ -{1\over 2} ~\sum_{i=1}^{N} {\vec \nabla}^2_i +{1\over 2} ~\sum_{i=1}^{N} 
{\vec r}_i^2 + \frac{g_1}{2} ~\sum_{\stackrel{i,j}{(i\ne j)}}^N 
\frac{{\vec r}_j^2}{X_{ij}^2} + \frac{g_2}{2} ~\sum_{\stackrel{i,j,k}{(i\ne 
j\ne k)}}^N \frac{{\vec r}_j \cdot {\vec r}_k}{X_{ij}X_{ik}} ~. 
\label{ham2} 
\eeq

We first obtain the exact many-body ground state of this Hamiltonian. As an 
ansatz for the ground state wave function, consider a solution of the form 
\beq
\Psi_0 (x_i,y_i)~=~ \prod_{i<j}^N ~X_{ij}^{\lambda} ~|X_{ij}|^{\alpha}~
\exp(-{1\over 2} \sum_{i=1}^N {\vec r}_i^2) ~.    
\label{ansatz}
\eeq
The ansatz for the ground state is bosonic (fermionic) for $\lambda =0 (1)$. 
Let us define 
\beq
g ~=~ \alpha ~+~ \lambda ~.
\eeq
Clearly $\Psi_0$ correctly incorporates the 
behavior of the true ground state in the asymptotic region, $|\vec r_i| 
\rightarrow \infty$ and for $g \ge 0$, $\Psi_0$ is regular. The 
eigenvalue equation now  takes the form 
\beq
H \Psi_0 ~=~[\frac{1}{2} (g_1-g(g-1)) \sum_{\stackrel{i,j}{(i\ne j)}}^{N} 
\frac{{\vec r}_j^2}{X_{ij}^2} + \frac{1}{2} (g_2-g^2) \sum_{\stackrel{i,j,k} 
{(i\ne j\ne k)}}^{N} \frac{{\vec r}_j \cdot {\vec r}_k}{X_{ij}X_{ik}} + 
gN(N-1) +N] \Psi_0 ~. 
\eeq
Therefore $\Psi_0$ is the exact many-body ground state for arbitrary number of 
particles of the Hamiltonian if 
\beq
g_1 ~=~ g(g-1),~~~ g_2 ~=~ g^2. 
\label{g12}
\eeq
Equivalently,
\beq
g ~=~ \frac{1\pm \sqrt{1+4g_1}}{2} ~,
\label{al}
\eeq
which immediately implies $g_1 \ge -1/4$ for physical states and $g_2$ is 
positive definite. Note 
that these relations are identical to the ones obtained in the CSM. While
$\Psi_0$ is a solution of the Schr$\ddot o$dinger equation in either of
the two branches given in (\ref{al}), the solutions in the lower branch
are regular only in the limited range $-1/4 \le g_1 \le 0$. The
solutions in the upper branch,however, are regular for all $g_1 
\ge  -1/4$. With this the ground state energy is given by
\beq
E_0 ~=~ N ~+~ gN(N-1).   
\label{egs}
\eeq
Note that this has exactly the form of the ground state energy of the CSM.

We emphasise that our objective here is not to find the general solutions for 
arbitrary 
$g_1$ and $g_2$, but to find a Hamiltonian whose solutions have the novel 
correlation in Eq. (\ref{xij}) built in. With 
the form of $g_1$ and $g_2$ given above in (\ref{g12}), the solutions found above 
are indeed the lowest energy states.  To see this \cite{proof}, let
\bea
Q_{x_i} ~&=&~ p_{x_i} ~-~ ix_i ~+~ i~g ~\sum_{j(j\ne i)}
\frac{y_j}{X_{ij}} ~, \nonumber \\
Q_{y_i} ~&=&~ p_{y_i} ~-~ iy_i ~-~ i~g ~\sum_{j(j\ne i)}
\frac{x_j}{X_{ij}} ~. 
\eea
It is easy to see that both $Q_{x_i}$ and $Q_{y_i}$ annihilate the ground 
state. The Hamiltonian can be recast in terms of these operators as
\beq
\frac{1}{2} ~\sum_i ~\left[ Q_{x_i}^{\dag} Q_{x_i} + Q_{y_i}^{\dag} Q_{y_i} 
\right] ~=~ H ~-~ E_0 ~, 
\eeq
where $E_0$ is given by Eq. (\ref{egs}). Clearly the operator on the left hand 
side is positive definite and annihilates the ground state wave function given 
by Eq. (\ref{ansatz}). Therefore $E_0$ must be the minimum energy that an 
eigenstate can have. 
Note that the total angular momentum $L = \sum_i (x_i p_{y_i} - y_i p_{x_i})$
commutes with the Hamiltonian, and may be used to label the states. 

While we have not been able to find the complete excited state spectrum
of the model, the eigenvalue equation for a general excited state may
be obtained as follows. From the asymptotic properties of the solutions of
the Hamiltonian in Eq. (\ref{ham2}), it is clear that $\Psi$ has the
general structure
\beq
\Psi (x_i,y_i) ~=~ \Psi_0 (x_i,y_i) ~\Phi(x_i,y_i) ~,
\eeq
where $\Psi_0$ is the ground state wave function, and $\Phi$ is a completely 
symmetric function of the coordinates. (Note that the bosonic or fermionic
nature of the eigenstates is completely specified by the choice of
$\lambda$). $\Phi$ satisfies the eigenvalue equation
\beq
[~-{1\over 2} ~\sum_{i=1}^{N}{\vec \nabla}_i^2 ~+~\sum_{i=1}^{N} {\vec r}_i 
\cdot {\vec \nabla}_i ~+~ g ~\sum_{\stackrel{i,j}{(i\ne j)}} \frac{1}{X_{ij}} 
(x_j \frac{\partial}{\partial y_i}-y_j \frac{\partial}{\partial x_i})~] ~\Phi ~
=~ (E - E_0) ~\Phi ~. 
\label{hamphi}
\eeq

We now consider a class of excited states built on this 
ground state, 
namely the radial excitations. These states are of the form
\beq
\Psi(\{x_i,y_i\})~=~\Psi_0(\{x_i,y_i\})\Phi(t); ~~~t =\sum_{i=1}^N r_i^2
\eeq
The eigenvalue equation for $\Psi$ can then
be easily reduced to a confluent hypergeometric 
equation\cite{Landau}, 
\beq
t \frac{d^2 \Phi(t)}{dt^2} + [b -t] \frac{d\Phi(t)}{dt}-a \Phi(t) = 0. 
\eeq
where, $b= E_0 = N+(\lambda+\alpha)N(N-1),~~ a=- {1\over 2}[E-E_0] $. The
admissible solutions are the regular confluent hypergeometric functions,
$\Phi(t)=M(a,b,t)$. The normalizability immediately imposes the restriction
$a=-n$, where $n$ is a positive integer and $\Phi(t)$ is in general a polynomial
of degree $n$. The corresponding eigenvalues follow immediately: 
$E = E_0+2n$. Once again these solutions have the
spectrum that is identical to that of states in the CSM. The tower structure of 
eigenvalues,
separated by a spacing of 2, is also reminiscent of what happens in the case of
anyons\cite{anyons}. Notice that until now we have only considered the completely
symmetric solutions modulo the correlation factor $X_{ij}$. Therefore the bosonic
or fermionic nature of the eigenstates is completely specified by the choice of
$\lambda$. 

We now seek exact solutions to Eq. (\ref{hamphi}) which are not in the 
same class as the simple solutions obtained above. We do this by eliminating 
the cross-terms involving the derivatives in Eq. (\ref{hamphi}). As in the 
case of the CSM \cite{S72}, we assume that such eigenstates have the form,
\beq
\Phi (x_i,y_i) ~=~ \sum_{m_i,n_i} a_{m_1\ldots m_N , n_1\ldots n_N}
\prod_{m_i,n_i}H_{m_i}(x_i)H_{n_i}(y_i) ~=~\sum_{m_i,n_i} 
a_{m_i,n_i}\prod_{m_i,n_i}H_{m_i}(x_i)H_{n_i}(y_i) ~, 
\eeq
where $m_i,n_i$ are positive integers with the restriction $\sum_i (m_i+n_i)= 
\cal N$, and the $H_m$'s are Hermite polynomials of degree $m$.
These are of course solutions of the non-interacting $N$-oscillator 
Hamiltonian. The expansion coefficients $a_{m_i , n_i}$ are 
determined by demanding that the cross-terms vanish when acting on the 
wavefunction above \cite{Kawakami}. Indeed the cross-terms vanish, provided
\bea
&4& (m_i+1)(n_j+1)a_{m_1\ldots m_{i+1}\ldots m_{j} \ldots n_{i} \ldots 
n_{j+1} \ldots} \nonumber \\
&+& (m_i+1) a_{m_1\ldots m_{i+1}\ldots m_{j} \ldots n_{i} \ldots n_{j-1} 
\ldots} + (n_j+1) a_{m_1\ldots m_{i-1}\ldots m_{j} \ldots n_{i} \ldots 
n_{j+1} \ldots} \nonumber \\
=&4& (m_j+1)(n_i+1)a_{m_1\ldots m_{i}\ldots m_{j+1} \ldots n_{i+1} \ldots 
n_{j} \ldots} \nonumber \\ 
&+& (m_j+1) a_{m_1\ldots m_{i}\ldots m_{j+1} \ldots n_{i-1} \ldots n_{j} 
\ldots} + (n_i+1) a_{m_1\ldots m_{i}\ldots m_{j-1} \ldots n_{i+1} \ldots 
n_{j} \ldots} ~, 
\eea
for all $ 0 \le m_i, n_i \le \cal N $. In general, it is difficult to 
obtain a closed-form solution for the coefficients $a_{m_i , n_i}$ by 
solving the above equation. However, in specific cases involving the low 
energy excitations, we have verified that it is possible to 
satisfy the above identity and the solutions exist. It is not apparent, 
however, that such solutions exist in general. If they do, then  
the energy spectrum of these excited states is given by $E_{\cal
N}= E_0 + \sum_i (m_i + n_i) = E_0+ \cal N$, where
$\cal N$ now labels the quanta of oscillator excitations above the ground
state. 

Finally, we point out the connection between our model Hamiltonian and the 
trigonometric Sutherland model \cite{S71}. Restricting the particles to 
move along the perimeter of a unit circle in the Hamiltonian (\ref{ham2}) 
without the confinement potential, we get 
\beq
H~=~ -{1\over 2} \sum_{i=1}^{N}{\partial^2\over \partial \theta_i} + 
\frac{g_1}{2} \sum_{\stackrel{i,j}{ (i\ne j)}}^{N} \frac{1}{\sin^2(\theta_i 
-\theta_j)} + \frac{g_2}{2} \sum_{\stackrel{i,j,k}{ (i\ne 
j\ne k})}^{N} [1+\cot(\theta_i-\theta_j)\cot(\theta_i -\theta_k)] ~, 
\label{ham5} 
\eeq
since $X_{ij}=\sin (\theta_i -\theta_j)$ now. Using the identity
\beq
\sum_{\stackrel{i,j,k}{ (i\ne j\ne k)}}^{N} 
\cot(\theta_i-\theta_j)\cot(\theta_i-\theta_k)] = -~ \frac{N(N-1)(N-2)}{3} ~,
\eeq
we immediately recover an analog of the trigonometric Sutherland model, but 
shifted by the constant $g_2~ N(N-1)(N-2)/3$. Note, however, that the  
potential in (\ref{ham5}) depends on the function $\sin(\theta_i 
-\theta_j)$,  rather than the chord-length which is proportional to 
$\sin[(\theta_i -\theta_j)/2]$ . 
 
In summary, we have constructed a nontrivial many-body Hamiltonian in two
dimensions with two- and three-body interactions which can be solved
exactly for the ground state and a class of excited states. The wave 
function involves a novel type of correlation which vanishes not only 
when the particle positions coincide, but also when the relative 
angle is a multiple of $\pi$. While the later may seem unphysical, the 
paucity of exactly solvable models even for the ground state in higher 
dimensions makes this an interesting system to study. The spectrum
displays a dramatic similarity to that of the Calogero-Sutherland model in one
dimension, namely, the ground state and the known excited states are all
shifted by exactly the same quantity proportional to $N(N-1)$, where $N$
is the number of particles. We also note that there are
many similarities between the spectrum of our Hamiltonian and the spectrum
of the many anyon Hamiltonian \cite{anyons}, without, however, the
complication introduced by the time-reversal violation in anyon physics. 

We are extremely grateful to Avinash Khare for many illuminating comments.
One of us (MVN) wishes to thank G. Date for the many techniques used in
this paper, and the Department of Physics and Astronomy, McMaster
University for hospitality. This work was supported by a grant from
Natural Sciences and Engineering Research Council of Canada.

\end{document}